 \documentclass[12pt,preprint]{aastex}

\shorttitle{Johnson et al.}
\shortauthors{Extragalactic UCHII Clusters}

\begin{document}

\def\HII  	{H\,{\small{\sc II}}}
\def\UDHII 	{UD{H\,{\small{\sc II}}}}
\def\UCHII      {UC{H\,{\small{\sc II}}}}
\def\etal 	{et~al.\/}

\title{A Sample of Clusters of Extragalactic Ultra Compact \HII\ Regions}

\author{Kelsey E. Johnson }
\affil{JILA, University of Colorado and National Institute for 
Standards and Technology; and Department of Astrophysical and 
Planetary Sciences, Boulder, CO 80309-0440}
\email{kjohnson@colorado.edu}

\author{Henry A. Kobulnicky}
\affil{Department of Astronomy, University of Wisconsin, 475 N. Charter St., 
Madison, WI 53706}
\email{chip@astro.wisc.edu}

\author{Philip Massey}
\affil{Lowell Observatory, 1400 Mars Hill Road, Flagstaff, AZ 86001 }
\email{massey@lowell.edu}

\and

\author{Peter S. Conti}
\affil{JILA, University of Colorado and National Institute for 
Standards and Technology; and Department of Astrophysical and 
Planetary Sciences, Boulder, CO 80309-0440}
\email{pconti@jila.colorado.edu}

\author{Draft of 30 May 2001}

\begin{abstract} We report on the detection of optically thick
free-free radio sources in the galaxies M33, NGC~253, and NGC~6946
using data in the literature.  We interpret these sources as being
young, embedded star birth regions, which are likely to be clusters of
ultracompact \HII\ regions.  All 35 of the sources presented in this
article have positive radio spectral indices ($\alpha > 0$ for $S_{\nu}
\propto \nu^{\alpha}$), suggesting an optically thick thermal
bremsstrahlung origin from the \HII\ region surrounding the hot stars.
The estimated emission measures for these sources are $EM_{6cm}
\gtrsim 10^8$ cm$^{-6}$ pc, and energy requirements indicate that the
sources in our sample have a range of a few to $\sim 560$ O7V star
equivalents powering their \HII\ regions.  Assuming a
Salpeter IMF with lower and upper mass cutoffs of 1 and 100 $M_\odot$,
respectively, this range in $N_{Lyc}$ corresponds to integrated
stellar masses of 0.1--60$\times 10^3 M_\odot$.  For roughly half of
the sources in our sample, there is no obvious optical counterpart,
giving further support for their deeply embedded nature; for most of
the remaining sources the correspondance to an optical source is
insecure due to relative astrometric uncertainty.  Their
luminosities and radio spectral energy distributions are consistent
with \HII\ regions modeled as spheres of plasma with electron
densities from $n_e \sim 1.5 \times 10^3$ to $n_e \sim 1.5 \times
10^4$ cm$^{-3}$ and radii of $\sim 1 - 7$ pc. Because of the high
densities required to fit the data, we suggest that the less luminous
of these sources are extragalactic ultracompact {H\,{\small{\sc II}}}
region complexes, those of intermediate luminosity are similar to W49
in the Galaxy, while the brightest will be counterparts to 30 Doradus
when they emerge from their birth material.  These objects constitute
the lower mass range of extragalactic ``ultradense {H\,{\small{\sc
II}}} regions'' which we argue are the youngest stages of massive star
cluster formation yet observed.  The sample presented in
this paper is beginning to fill in the continuum of objects between
small associations of ultracompact HII regions and the analogous massive 
extragalactic clusters that may evolve into globular clusters.

\end{abstract}


\section{INTRODUCTION} 
In relatively nearby starburst galaxies, recent star formation
activity has typically been resolved into massive star clusters.
While old globular clusters are ubiquitous at the current epoch, only
over the past decade have we begun to find their younger and bluer
siblings, ``super star clusters" (SSCs), in significant numbers
\citep{whitmore00}.  The large number of globular clusters in the
local universe appear to have been formed during the early stages of
galaxy evolution. Therefore, studying the genesis of local SSCs can
yield information about the environment in which globular clusters
formed in the early universe.  The conditions required for massive
star cluster formation are far from understood, but it seems clear
that extreme environments, uncommon in the local universe, are
necessary.  Current theories suggest that high pressures, such as
those due to large-scale shocks in merging galaxies
\citep{elmegreen97}, are required in order to form bound massive star
clusters.  In fact, there is a large body of evidence that globular
clusters are formed in galactic interactions and mergers
\citep[e.g.,][]{whitmore00}.

Recently, a handful of very young massive star clusters still embedded
in their birth material have been discovered in the galaxies NGC~5253
\citep{turner98}, He~2-10 \citep{kj99}, NGC~2146 \citep{tarchi00}, and
NGC~4214 \citep{beck00}.  Embedded in these heavily enshrouded
clusters are hundreds of young massive stars that create surrounding
{H\,{\small{\sc II}}} regions and manifest themselves as optically
thick free-free radio sources.  Similar dense, inverted-spectrum
{H\,{\small{\sc II}}} regions exist around individual stars in our
galaxy (i.e., ultracompact {H\,{\small{\sc II}}} regions,
UC{H\,{\small{\sc II}}}s, Wood \& Churchwell 1989), albeit on a vastly
smaller scale. Because of their apparent spectral similarity to
galactic UC{H\,{\small{\sc II}}}s, \citet{kj99} dubbed these
extragalactic objects ``ultradense {H\,{\small{\sc II}}} regions''
(UD{H\,{\small{\sc II}}}s).  The physical properties of these
clusters are truly remarkable; the estimated sizes (a few parsecs),
stellar masses (a few $\times 10^{5-6}$ $\rm M_{\odot}$), ionizing
luminosities ($\rm N_{Lyc}\sim 10^{51-53}$ erg), and ages (possibly as
young as a few $\times 10^5$ years) of the newly discovered
UD{H\,{\small{\sc II}}}s imply that we may be witnessing the birth
process of super star clusters.  The discovery of these
UD{H\,{\small{\sc II}}}s allows us to begin observing the earliest
stages of the massive star cluster formation for the first time.

Because UD{H\,{\small{\sc II}}}s have only recently been identified in
the literature, one of our primary goals is simply to expand the
sample of known objects in order to understand their properties in a
statistical sense.  To this end, we have searched for previously
published radio observations that may have also serendipitously
detected UD{H\,{\small{\sc II}}}s.  Multiwavelength radio observations
taken to study supernovae remnants in nearby galaxies are particularly
useful for this purpose.  In most cases, the original authors have
identified the candidate UD{H\,{\small{\sc II}}}s we present here as
``thermal sources'' or ``{H\,{\small{\sc II}}} regions.''  We seek
sources with positive radio spectral indices ($\alpha > 0$, where
$S_{\nu} \propto \nu^{\alpha}$), implying an optically thick free-free
origin.  In this article, we utilize previously published radio
observations of M33, NGC~253, and NGC~6946 to identify new candidate
UD{H\,{\small{\sc II}}}s, determine their number of ionizing stars,
and estimate their physical sizes, densities, and masses.

\section{GALAXIES IN THIS SAMPLE}

Here we overview basic properties of the galaxies in our sample and
only highlight the radio observations utilized in this paper.  We
refer the reader to the original papers for more detail on the
observations and reduction.  Because the observations were compiled
from a variety of sources, the observations are heterogeneous in
nature, and in some cases, non-ideal for our purposes.  For example,
in order to obtain accurate relative fluxes at multiple frequencies it
is desirable to have well matched beam sizes; we will discuss the
impact of mismatched beams in Section~\ref{detection}.  However, these
data provide a valuable resource for this fledgling field and will
serve as a useful baseline for follow-up observations.

\subsection{\it M33} M33 (NGC~598) is a Local Group spiral galaxy 
(Sc(s)II-III), nearly face-on \citep[$i = 55^{\circ}$,][]{garcia91},
and it is rich in {H\,{\small{\sc II}}} regions \citep{hodge99,
courtes87, boulesteix74}.  Young ($< 10^8$ yr) clusters in M33 were
recently studied by \citet{chandar99}, and were found to have masses
of $6 \times 10^2 - 2 \times 10^4 M_\odot$, smaller than that typical
for its old globular clusters.  M33 has a star formation rate of
log(SFR)~$= -2.47$~M$_\odot$~yr$^{-1}$~kpc$^{-2}$ \citep{kennicutt98},
making it the least prolific galaxy in this sample.  At a distance of
840 kpc \citep{freedman91}, it is also the closest of the galaxies in
this paper.

The radio observations of M33 that we use were originally obtained by
\citet{duric93} and also presented by \citet{gordon99}.  Both Very
Large Array (VLA) B configuration and Westerbork Synthesis Radio
Telescope (WSRT) observations were made of M33 at 6 and 20~cm (4.84
and 1.42~Ghz) in several pointings to obtain good coverage of the
galaxy.  The primary beams at 20~cm were $30'$ with the VLA and $36'$
with WSRT.  At 6~cm both telescopes had primary beams of $9'$.  The
6~cm observations had resolutions ranging from $5''$ to $10''$, and
the 20~cm observations had resolutions ranging from $5''$ to $15''$,
depending on the uv taper used in the cleaning process.  Matched beam
observations were convolved to identical beams of $7''$ FWHM.  All
images have an average 3$\sigma$ rms noise level of $\approx 150~\mu$Jy
per beam.  At the adopted distance of 840 kpc \citep{freedman91},
1~Jy$ = 8.5 \times 10^{26}$~erg~s$^{-1}$~Hz$^{-1}$.

\subsection{\it NGC~253}
At 2.5 Mpc \citep{turner85}, NGC~253 is a member of the Sculptor
Group.  This galaxy is a nearly edge-on spiral galaxy (Sc(s)), and
contains a radio continuum plume emanating from the central starburst
region rising perpendicular to the disk \citep[e.g.,][]{carilli92}.
NGC~253 has the highest star formation rate of the galaxies in this
sample with log(SFR)~$ = 1.24$~M$_\odot$~yr$^{-1}$~kpc$^{-2}$
\citep{kennicutt98}.  Four young SSCs were discovered optically in the
central region of NGC~253 by \citet{watson96}; however, only one of
these clusters is apparent in the mid-IR observations of
\citet{keto99}, who propose that it might be exceptionally young.
\citet{ulvestad97} also discuss the nature of the brightest thermal
radio source in NGC~253, a point that we will revisit in
Section~\ref{results}.

NGC~253 was observed by \citet{ulvestad97} with the B configuration of
the VLA at 1.3, 2, 3.6, 6, and 20~cm (23.56, 14.94, 8.44, 4.86, and
1.49~Ghz) between 1987 and 1995.  Additional observations at 1.3 and
2~cm were made using the A configuration.  As the beam sizes at 6 and
20 cm are significantly larger than at 1.3, 2, and 3.6 cm, we have not
included this longer wavelength data in our analysis because of the
unwanted background contribution from non-thermal emission.  The
duration of the observations in the B-array were 4.1, 5.6, and 7.0
hours at 1.3, 2, and 3.6~cm, respectively.  The A-array observations
at 1.3 and 2~cm were 2.8 and 4.2 hours, respectively.  The primary
beams varied from $2'$ to $5'$ (from 1.3~cm to 3.6~cm, respectively) and
easily contain the area of interest.  The synthesized beams ranged
from $0.13'' \times 0.07''$ to $0.42'' \times 0.24''$ at 1.3~cm,
$0.20'' \times 0.10''$ to $0.59'' \times 0.35''$ at 2~cm, and $0.33''
\times 0.19''$ to $0.36'' \times 0.21''$ at 3.6~cm.
\citet{ulvestad97} estimate the uncertainties in these data as $\sim
20$\% for 1.3 cm, and $\sim 10$\% for the 2~cm and 3.6~cm data.  The
3$\sigma$ rms noise levels are $\approx$~660, 240, and 150~$\mu$Jy,
respectively, for the 1.3~cm, 2~cm, and 3.6~cm data.  At the adopted
distance of 2.5 Mpc \citep{turner85}, 1~Jy$ = 7.5 \times
10^{27}$~erg~s$^{-1}$~Hz$^{-1}$.

\subsection{\it NGC~6946} 

NGC~6946 is the most distant galaxy in this sample at 5.1~Mpc
\footnote{Distances as high as 10.1~Mpc have been estimated
\citep[e.g.,][]{sandage81}.} \citep{devaucouleurs79}, and is a spiral
galaxy of type Sc(s)II with an inclination angle of $i=34^{\circ}$
\citep{garcia91}.  The star formation rate in NGC~6946 is log(SFR)~$=
-1.88$~M$_\odot$~yr$^{-1}$~kpc$^{-2}$ \citep{kennicutt98}.
\citet{larsen99} have detected 107 young massive star clusters, one of
which has received particular attention because of its age ($\sim 15$
Myr) and mass ($\sim 5 \times 10^5 M_\odot$) \citep{elmegreen00}.
Sub-mm images obtained by \citet{bianchi00} show 850 $\mu$m emission
which, in some cases, may be associated with radio sources discussed
in this paper.

The VLA was used to survey NGC~6946 with the B and A arrays at 6 and
20~cm (4.86 and 1.45~Ghz), respectively, by \citet{lacey97}.
Approximately 8.6~hours were spent in at each frequency.  The primary
beams were $9'$ and $30'$ at 6~cm and 20~cm, respectively.  The
resulting beam sizes at the two wavelengths are very well matched:
$1''.9 \times 1''.6$ at 6 cm and $1''.9 \times 1''.7$ at 20 cm.  Flux
uncertainties are typically $\approx 20$\%, and the 3$\sigma$ rms
noise levels are $\approx$~60 and 50 $\mu$Jy for the 20 and 6~cm data,
respectively.  At the adopted distance of 5.1 Mpc
\citep{devaucouleurs76}, 1 Jy $= 3.1 \times
10^{28}$~erg~s$^{-1}$~Hz$^{-1}$.  These observations have also been
analyzed by \citet{hyman00}, who derive a luminosity function for
{H\,{\small{\sc II}}} region candidates with a power-law index at 6~cm
of $\approx -1.2$, but exclude sources with spectral indices of
$\alpha > 0.2$.  As result, the most optically thick sources were
excluded from their sample.  Never the less, the seven sources
presented in \citet{hyman00} with $\alpha > 0$ are also included in
this paper (our sources 1, 2, 4, 6, 7, 8, and 10 in Tables~\ref{tbl-3}
and \ref{N6946.tab}).

\section{RESULTS \label{results}} 

\subsection{\it Detection of UD{H\,{\small{\sc II}}} Candidates 
\label{detection}} 
Our primary tool for distinguishing UD{H\,{\small{\sc II}}}s from
other types of radio sources, such as supernovae remnants (SNR), is
the radio spectral index $\alpha$ (where $S_{\nu} \propto
\nu^{\alpha}$).  While optically thin sources have $\alpha = -0.1$,
optically thick sources have $\alpha > -0.1$ ($\alpha$ approaches 2 in
the black body limit), and typical non-thermal objects have $\alpha <
-0.4$.  Unfortunately, not all SNRs obey this convention --- in
particular, there is a class of ``composite'' or ``plerionic'' SNRs
\citep[as coined by][]{weiler78} that are known to have atypically
flat spectral indices ($-0.3 < \alpha < 0.0$).  In fact, in the first
few hundred days after a supernova explosion, $\alpha$ can actually be
positive \citep{weiler86}.  However, given this extremely short
timescale, we would have to be extremely fortunate (or unfortunate) to
have one of these objects contained in our sample.  A second set of
observations taken with adequate time separation, such as those
employed by \citet{kj99} would resolve this issue with confidence.

Consequently, in order to select UD{H\,{\small{\sc II}}} candidates
from the radio data, we have applied the condition that these sources
must have radio spectral indices $\alpha > 0$ (higher fluxes at
shorter wavelengths).  This criterion should exclude most other
classes of radio-emitting objects, but it may also exclude genuine
UD{H\,{\small{\sc II}}} regions depending upon the exact instrumental
angular resolution and frequencies observed.  For example,
low-frequency observations of optically thick thermal sources could be
contaminated by steep-spectrum non-thermal sources, such as SNRs, and
thus fail the $\alpha > 0$ test, especially at longer wavelengths.
Furthermore, more distant sources, or sources observed using beam sizes
larger than the thermal emitting region, may be contaminated by
steep-spectrum, non-thermal background emission that dominates the
disks and halos of most galaxies.  Both of these effects will act to
disguise an optically thick thermal bremsstrahlung signature.
Thermal sources with the lowest emission measure, and thus the lowest
free-free optical depth will be affected most strongly.  Sources with
the highest emission measures are more likely to maintain a positive
spectral signature and be correctly identified.  Fortunately, it is
precisely these sources with high emission measures that are most
likely to host ultra-young stellar clusters.

Because we are relying on relative fluxes at different frequencies to
identify {\UDHII}s, it is critical that we understand the impact of
beam sizes on these flux measurements.  In some cases (as for
NGC~6946), the multifrequency observations were done in different
array configurations to match beam sizes at different frequencies.
However, in some observations presented here the beam sizes are not
well matched (as for M33).  In these cases, typically a single
configuration is used for all frequencies causing the beam size at the
higher frequency to be smaller than the beam size at lower frequency.  If
the object ({\UDHII}) is point-like with respect to the beams, the
larger (lower frequency) beam will contain more non-thermal background
contribution (as discussed above), which can only cause these objects
to appear {\it less} thermal.  If the object is extended, the lower
frequency beam will contain more flux from the same physical
environment than the higher frequency beam --- again, artificially
boosting the relative flux at the lower frequency, and also causing
the object to appear {\it less} thermal.  In other words, if the
beam size at lower frequencies is larger than the beam size at higher
frequencies (as is the case for the mis-matched beams in this paper),
it will only disguise the optically thick signature of {\UDHII}s and
{\it not} artificially imitate an optically thick thermal spectral index.

The condition that UD{H\,{\small{\sc II}}} candidates must have
positive spectral indices has resulted in the detection of fourteen
sources in M33, five sources in NGC~253, and sixteen sources in
NGC~6946.  These sources are listed in Tables~\ref{tbl-1} --
\ref{tbl-3} along with their spectral indices ($\alpha$, where
$S_{\nu} \propto \nu^{\alpha}$).  Within the uncertainties, eight of
these objects are not inconsistent with having $\alpha \lesssim 0$,
and we cannot rule out an optically thin \HII\ region as the source.
The remaining twenty-seven objects are only consistent with having an
optically thick origin, and we consider these sources to be strong
candidates for having an {\UCHII}-like origin.
\citet{ulvestad97} have previously discussed the brightest of these
five sources in NGC~253 (source \#3 in this paper).  Their derived
properties ($Q_{Lyc} = 5.2 \times 10^{51}$~s$^{-1}$, size $=2.4 \times
1.2$~pc, $EM = 2 \times 10^8$~cm$^{-6}$~pc, and $n_e = 1.3 \times
10^4$~cm$^{-3}$) are in excellent agreement with the values we derive
for this object in this section.  However, the properties of the other
four sources were not discussed, and we wish to add these sources to
the current sample of {\UDHII}s in the literature.

\subsection{\it Comparison to Optical Images} 
As the gas and dust associated with {\UDHII}s dissipates and the
extinction lessens, these objects will become more easily visible in
optical light.  On the other hand, {\UDHII}s will become less obvious
in radio observations as their ambient densities decrease and their
thermal bremsstrahlung emission decreases as a result.  {\it
Consequently, optical surveys of {\HII} regions will be biased against
detecting the youngest {\HII} regions and radio surveys will tend to miss
the older {\HII} regions.}

Therefore, in order to understand how these UD{H\,{\small{\sc II}}}
candidates are related to star formation, past or present, it is
useful to examine optical images near the source positions.  In most
cases, this comparison reveals whether the \UDHII\ regions are
associated with star formation visible in optical light, diffuse
{H\,{\small{\sc II}}} emission, or are completely obscured in this
wavelength regime.  In making such comparisons in crowded fields, we
must keep in mind that the radio positions are no better than about
half a beam width, e.g., $3-4''$ in M33, $1''$ in NGC~6946, and
sub-arcsecond in NGC~253.  To make these comparisons, we have tied
the coordinate system of the optical images to the {\it HST} Guide
Star ``system.'' This leads to small internal errors ($0''.3$ RMS)
within a galaxy, but it is well known that the same star will have
coordinates that may differ by $2''$ or more from plate to plate in the
Guide Star Catalog. We thus expect systematic offsets between the
optical and radio positions by as much as $3''$.  This is possibly an
overestimate, but we cannot rule out the physical correspondence of
optical and radio sources within the astrometric uncertainty.

The locations of the UD{H\,{\small{\sc II}}} regions in M33 were
compared with B-band images from the 0.9 m telescope on Kitt Peak; see
\citet{massey96} for a complete description of the observations.  We
also searched the catalogs of {H\,{\small{\sc II}}} regions in M33
given by \citet{boulesteix74}, \citet{courtes87}, \citet{hodge99}.
M33 is rich in {H\,{\small{\sc II}}} regions, and all of the
UD{H\,{\small{\sc II}}}s are possibly associated with previously known
{H\,{\small{\sc II}}} regions, although in some cases the connection
is insecure (Table~\ref{M33.tab}).  However, only about half of the
UD{H\,{\small{\sc II}}}s appear to be associated with stellar light
apparent in the B-band image.  The locations of the detected
UD{H\,{\small{\sc II}}}s with respect to the B-band image are shown in
Figure~\ref{M33bband}.
\notetoeditor{There are twelve images which are part of this 
figure.  If possible, I would like them all on a single page, 
row1: abc, row2: def, row3:ghi, row 4: jkl}

The radio observations of NGC~253 were compared to archival F656N
(narrow-band H$\alpha$) and F814W (I-band) Hubble Space Telescope
images (Figure~\ref{N253}).  \notetoeditor{If possible, I would like
the two parts of this figure displayed side-by-side.}  While there is
diffuse emission in the vicinity of all five UD{H\,{\small{\sc II}}}s
in this galaxy, only one of them (source \#2) is clearly identified
with a compact optical object.  These results are presented in
Table~\ref{N253.tab}.

For NGC~6946, we compare the locations of the thermal radio sources
with optical R-band and H$\alpha$ images previously published by
\citet{larsen99} (Figure~\ref{N6946}).  \notetoeditor{There are five
pairs of images in this figure (pairs: ab, cd, ef, gh, ij).  The two
members of each pair should be displayed side-by-side.}  However,
because of the large angular size of the galaxy ($\sim 10' \times
10'$) this image does not have a wide enough field to contain all of
the radio sources.  Therefore, we also use images from the the STScI
Digitized Sky Survey
\footnote{The compressed files of the Space Telescope Science
Institute Quick-Survey of the northern sky are based on scans of
plates obtained by the Palomar Observatory using the Oschin Schmidt
Telescope.} for comparison in these cases (Figure~\ref{N6946dss}).
\notetoeditor{The two parts of this figure should be displayed
side-by-side.}
About half of the sources in NGC~6946 have possible optical
counterparts or diffuse emission, and the remaining half have neither.

While the sources for which there are no obvious optical counterparts
must be deeply enshrouded, and therefore are likely to be extremely
young, it isn't clear whether detection of light in the optical regime
rules out extreme youth for UD{H\,{\small{\sc II}}} regions.  We make
this tentative statement for several reasons: (1) in most cases
presented in this sample, the identification of an optical counterpart
is insecure due to the possible systematics between the optical and
radio positions, as well as the relatively large synthesized radio
beam-width for M33; (2) in cases where individual stars are resolved
(M33), the {\it number} of stars required to create the Lyman
continuum flux are not apparent in the optical images, suggesting
either a misidentification due to pointing uncertainty, or a number of
the individual stars are, in fact, still enshrouded.  This scenario
has actually been observed in the galactic \UCHII\ complex W49A where
{\it some} of the stars in the complex appear to have emerged from
their birth cocoons while the rest of the complex remains enshrouded
\citep{conti01} ; and (3) if the dominant source of opacity is
Thompson scattering (and not dust), a source with a radius of 5 pc
could have electron densities as high as $n_e = 10^5$~cm$^{-3}$ and
still have opacities as low as $\tau \sim 1$.  Therefore, we conclude
that the possible identification of optical counterparts does not
preclude the youth of these objects.  However, the sources for which
there are no optical counterparts are more likely deeply embedded in
their natal molecular clouds, and therefore extremely young.

\subsection{\it Modeled Properties} 

Given the luminosities and radio spectral energy distributions of
\HII\ regions, their physical parameters such as size and electron
density can be estimated.  Using the analytical approximation of
\citet{mezgerhenderson67}, we can estimate the emission measure,
$EM=\int n_e^2 dl$, given an electron temperature, $T$, the observing
frequency, $\nu$, and the optical depth at that frequency.

\begin{equation}
EM({\rm cm}^{-6}{\rm pc}) = 12.2 \left[\frac{T_e}{({\rm K})}\right]^{1.35}
\left[\frac{\nu}{({\rm Ghz})}\right]^{2.1}\tau . 
\end{equation} 

The positive spectral index for these sources arises from free-free
emission where $\tau \gtrsim 1$, therefore we assume $\tau = 1$ as a
lower limit.  The electron temperature of Galactic {\UCHII}s is
typically $T_e = 8000 \pm 1000$~K \citep[e.g.,][]{afflerbach96}, which
we adopt for this estimate.  The resulting emission measures for each
of the wavelengths used in this sample range from $\sim 0.05 - 16
\times 10^8$~cm$^{-6}$~pc (for $\tau > 1$ these emission measures will
be correspondingly higher).  Although the radio maps constrain the
size of the emitting region only weakly~\footnote{ The radio maps
constrain the sizes of the emitting regions to diameters $\leq 4$ pc
in the case of NGC~253, $\leq 28$ pc in the case of M~33, and $<50$ pc
in the case of NGC~6946}, it is clear that electron densities in
excess of 1000~cm$^{-3}$ are required to produce the observed emission
measures.  For comparison, typical giant \HII\ regions observed
optically have electron densities $\approx 10^2$~cm$^{-3}$
\citep{kennicutt84}.

In order to better constrain the properties of the \UDHII\ regions,
following \citet{kj99} we have modeled \HII\ regions as homogeneous
spheres of plasma with uniform electron density and temperatures of
8000 K.  Only free-free emission and absorption processes are
considered.  Varying the radius, $R$, and electron density, $n_e$, we
modeled the radio spectral energy distribution resulting from thermal
bremsstrahlung emission and self-absorption.  These results are
illustrated in Figures~\ref{M33.plot}-\ref{N6946.plot} along with the
data from Tables~1--3.  Figure~5 shows the 20 cm and 6 cm observations
of \UDHII\ regions in M33 along with two sets of model \HII\ regions
spectral energy distributions.  Different symbols distinguish the
\UDHII\ candidates from Table~1.  Solid lines show \HII\ region models
with electron densities of $n_e=5000$~cm$^{-3}$ and radii of 0.7~pc
and 1.9~pc.  Although the modeled luminosities closely match the
observed luminosities, the spectral index of these models is generally
steeper than the data.  A second class of models with electron
densities of $n_e=1500$~cm$^{-3}$ and radii of 1.5~pc to 3.0~pc more
closely match the observed spectral indices and luminosities.  In the
case of NGC~6946, densities of 1500~cm$^{-3}$ to 5000~cm$^{-3}$ fit
the 20~cm and 6~cm data for radii ranging from 2~pc to 7~pc (only a
representative range of data from Table~3 are plotted).  In the case
of NGC~253, observations reveal that the \UDHII\ regions are optically
thick at frequencies as high as 15 GHz.  Figure~\ref{N253.plot}
illustrates that electron densities between 10,000~cm$^{-3}$ and
15,000~cm$^{-3}$ are required to fit the data.  Not only are the
\UDHII\ regions in NGC~253 intrinsically more luminous than in the
other two galaxies, they also have higher densities.

It is reasonable to ask whether the simple homogeneous sphere models
presented here are sufficient to infer the physical characteristics
(sizes, densities) of the {\UDHII}s.  Realistically, we expect that
the radio sources identified as {\UDHII}s are not simply monolithic
dense \HII\ regions.  Rather, we expect that they are collections of
several hundred ultra-compact \HII\ regions ($n_e=10^5$~cm$^{-3},
R=0.1$~pc) embedded in a more tenuous inter-\UCHII\ medium (see
\ref{W49A}).  However, two galaxies, NGC 253 (Ulvestad \& Antonucci
1997) and NGC 5253 (Turner \etal\ 1999) are near enough and have
observations with high enough resolution that the radio imaging can
directly constrain the sizes of the emitting regions without recourse
to model assumptions.  In both cases, our simple homogeneous spherical
models described above produce size estimates in excellent agreement
with the high-resolution radio maps.  The deconvolved diameters of the
radio sources in NGC 253 are 2-4 pc, in excellent agreement with the
best-fitting models shown in Figure~\ref{N253.plot}.  In NGC 5253, the
single \UDHII\ radio source has a deconvolved diameter of 1-2 pc
\citep{turner00}, consistent with the expectations of the simple model
predictions based on the observed radio luminosities of Turner \etal\
(1998).  As further support for our simple models, recently
\citet{mohan01} have made slightly more sophisticated multi-density
models for radio recombination line observations of He~2-10 and and
NGC~5253; their results are in excellent agreement with the results we
find with our simple models.

Given the agreement in these two nearby cases, we believe that the
simple two-parameter models are sufficiently instructive to make
meaningful inferences about the sizes and densities of {\UDHII}s based
on radio continuum luminosities.  More realistic models would include
an arbitrary number of \UCHII\ regions within each \UDHII\, each
with its own density profile, and an inter-\UCHII\ medium with a
specified density profile and temperature distribution.  However,
relaxing the simple two-parameter approach results in the number of free
parameters and computational complexity growing rapidly.  

\section{DISCUSSION}

\subsection{\it Stellar Content}

The production rate of Lyman continuum photons,
and, thus, the stellar content of each \UDHII\ region
can be estimated from the thermal radio luminosity
following \citet{condon92},
\begin{equation}
\left( \frac{Q_{Lyc}}{{\rm s}^{-1}}\right) \ge 6.3 \times 10^{52} 
\left( \frac{T_e}{10^4{\rm K}} \right)^{-0.45}
\left( \frac{\nu}{{\rm GHz}}\right)^{0.1}
\left( \frac{L_{thermal}}{10^{27} {\rm erg~s}^{-1} {\rm Hz}^{-1}}\right).
\end{equation}
Since the non-thermal component becomes weaker at higher frequencies,
for each of the {\UDHII}s in this sample we use the
luminosity measured at the highest frequency in each data set to
determine $Q_{Lyc}$.  The resulting values determined from this method
are presented in Tables~\ref{tbl-1}~--~\ref{tbl-3}.  One should also
bear in mind that the $Q_{Lyc}$ values determined with this method
could suffer from two different problems: (1) if the source is, in
fact, optically thick even at the highest frequencies measured, this
method will {\it underestimate} the actual ionizing luminosity, and (2)
if there is contamination from background non-thermal emission in the
beam at the frequency used, the ionizing luminosity will be {\it
overestimated}.  While these two issues have an opposite affect on
$Q_{Lyc}$, we can not determine the magnitude of either with the data
presently available.

The production rate of Lyman continuum photons from these sources can
be used to estimate the number of massive stars powering the observed
emission.  Following the convention of \citet{vacca94}, a ``typical''
O-star (type O7V) produces $Q_{Lyc} = 1.0 \times 10^{49}~s^{-1}$.
Therefore, the $Q_{Lyc}$ values in Tables~\ref{tbl-1}-\ref{tbl-3} can
be directly translated into the number of ``equivalent'' O7V stars
(O7V*).  Using this method, we see that the UD{H\,{\small{\sc II}}}
regions in this sample have $\approx 3 - 46$ O7V* stars in M33,
$\approx 60 - 560$ in NGC~253, and $\approx 10 - 360$ in NGC~6946.
Moreover, the {\it total} stellar mass of these objects can be
estimated using the Starburst99 models of \citet{leitherer99} with
solar metallicity, Salpeter IMF, and lower and upper mass cutoffs of
$1 M_\odot$ and $100 M_\odot$, respectively (decreasing the lower mass
limit will increase the total stellar mass estimate).  Using these
parameters, star clusters producing this range of Lyman continuum
photons at ages $\sim 0 - 5$~Myr would have total masses of $\sim
100-60000 M_\odot$.

These $Q_{Lyc}$ and mass values are smaller than those found for the
UD{H\,{\small{\sc II}}}s in NGC~5253 \citep{turner98}, Henize~2-10
\citep{kj99}, and NGC~2146 \citep{tarchi00}, which have $> 750$ O7V*
stars and masses $> 10^5 M_\odot$.  This is not a surprising result --
it is likely that we will only find massive star clusters forming in
intense starburst events; \citet{elmegreen97} have demonstrated the
need for extremely high-pressure environments, such as those found in
starburst galaxies, to produce bound massive star clusters.  However,
in less formidable environments, we should expect to find a continuous
range of {\UDHII}s --- from single {\UCHII}s and \UCHII\ complexes
(such as those found in the Galaxy), to the massive bound clusters
(such as those found in NGC~5253, Henize~2-10, and NGC~2146) that may
evolve into globular clusters.  The sample presented in this article is
beginning to fill in this continuum of objects.

\subsection{\it Comparison to W49A \label{W49A}} 

One of the most well studied \UCHII\ complexes in the
Galaxy is W49A, which makes it well suited for comparison to the
\UDHII\ regions in this sample.  First detected by
Westerhout (1958) in his radio survey, W49A has since been resolved
into at least 30 \UCHII\ regions \citep{depree97}, and is
estimated to have $\sim 100$ O7V* stars \citep{smith78,vacca94}.  It
also appear that a few of the stars in W49A have begun to emerge from
their birth cocoons while the rest of the cluster remains deeply
enshrouded \citep{conti01}.  W49A is $\sim 13$ pc in diameter
\citep{depree97}, and would fit within the beam sizes utilized in this
paper if located at the same distances as the galaxies in this sample;
W49A would have angular sizes of $\sim$ $2''.9$, $1''.2$, and $0''.6$,
respectively, in M33, NGC~253, and NGC~6946.  Because W49A is
significantly closer \citep[$D = 11.4$~kpc;][] {gwinn92}, than the
galaxies presented in this paper, any observations we use for
comparison need to be at sufficiently low resolution as to include the
entire W49A region.  This condition is well-satisfied by the
pioneering observations of \citet{mezger67}, who obtained $\sim 3 -
4'$ resolution radio maps of W49A at several wavelengths.

In Figure~\ref{W49A.plot}, we present the radio spectral energy
distribution of W49A \citep{mezger67} in comparison to the {\it mean}
\UDHII\ region from M33, NGC~253, and NGC~6946.  It is clear that the
integrated radio spectral energy distribution of W49A is almost
identical to those of the \UDHII\ regions in this sample.
Figure~\ref{histograms} \notetoeditor{There are three parts to this
figure.  If possible, I would like them displayed in a vertical stack
in the same column, with 'a' on top, and 'c' on the bottom.} shows
histrograms of the 6~cm luminosities for the sources in M33 and
NGC~6946 and the 3.6~cm luminosities (6~cm data is not included in
this sample) for NGC~253 along with an arrow indicating the luminosity
of W49A.  Comparing the luminosity of W49A to those of the \UDHII\
regions in M33, it is clear that W49A is up to 10 times {\it more}
luminous than the M33 sources.  In NGC~6946 W49A would be one of the
more luminous \HII\ regions.  However, in the case of NGC~253, W49A
would be one of the {\it least} luminous objects --- a magnitude or
more fainter than the most luminous \UDHII\ regions detected in this
galaxy.

The dotted line in Figure~\ref{histograms}a illustrates the luminosity
function of optically selected \HII\ regions in M33 \citep{smith89}
who find $N(L) \propto L^{-2.3} dl$.  This comparison shows that the
luminosity function of \UDHII\ regions is consistent with the normal
\HII\ region luminosity function, suggesting that the {\UDHII}s are
simply a phase in the formation of many, perhaps most, \HII\ regions.
Furthermore, if the formation of massive star clusters is primarily
related to the intensity of the star formation a galaxy is currently
undergoing, it is not surprising that the \UDHII\
regions in M33 and NGC~6946 are similar to the \UCHII\
complexes in the Milky Way which has a similar star formation rate.
NGC~253, by contrast, has a much higher star formation rate based on
its FIR luminosity, although high visual extinctions make the
$H\alpha$ luminosities very uncertain.  It certainly contains a more
intense starburst, which is also in accord with the more massive
UD{H\,{\small{\sc II}}} regions it hosts.

\subsection{\it On the Youth of \UDHII\ Regions}
It seems likely that the fraction of time a super star cluster spends
in the \UDHII\ phase is a small fraction of the massive star lifetime,
perhaps 10-15\%, in accord with the estimated lifetimes of individual
{\UCHII} regions based on the number of \UCHII\ regions compared with
the number of optically visible O stars in the Galaxy
\citep[e.g.,][]{wood89b}.  Indeed, \UCHII\ region lifetimes have been
a topic of much discussion since \citet{wood89a} introduced the
``lifetime problem''; in short, if \UCHII\ regions are significantly
overpressured with respect to the surrounding ISM, they should expand
and dissipate on time scales $\approx 10^4$~years.  However, the number
of {\UCHII} regions observed is greater than is allowed for by this
time scale.  Several mechanisms have been proposed to address this
issue, most of which are likely to also be applicable to {\UDHII}s.
\citet{wood89a} proposed that infalling matter or bow shocks might act
to increase the external pressure, thus extending the \UCHII\ phase.
It is also possible that the ambient pressure is typically
significantly higher than the value used by \citet{wood89a} as proposed
by \citet{depree95}.  The lifetimes of \UCHII\ regions could also be
extended if they are replenished by material photoevaporated from the
surrounding circumstellar disks \citep[e.g.,][]{hollenbach94}.

The first argument for the extreme youth of \UDHII\ regions is simply
by analogy to \UCHII\ regions in the Galaxy.  If {\UDHII}s are
composed of individual \UCHII\ regions, we should expect them to have
similar lifetimes provided that star formation is relatively
instantaneous over the massive star cluster.  As is the case for
\UCHII\ regions, the densities in {\UDHII}s are extremely high,
and the implied pressures constitute an over-pressure compared to
typical ISM pressures.  To first order, such over-pressed regions must
expand and disperse on time scales comparable to the sound-crossing
time scale which is a few$\times10^5$ yr (see Kobulnicky \& Johnson
1999 for more complete details).  However, as discussed above for
\UCHII\ regions, it is entirely possible that the ambient pressure
around {\UDHII}s is not ``typical'' of the global ISM.

The second piece of evidence suggesting extreme youth is the fraction
of ionizing stars in \UDHII\ regions compared to the fraction of
ionizing stars in conventional \HII\ regions.  For M33 and NGC~6946
the minimum implied total Lyman continuum photon production rate, $Q$,
of the \UDHII\ regions is $1.9\times10^{51}~s^{-1}$ and
$10\times10^{51}~s^{-1}$ respectively (provided there is no leakage
from the enshrouding cocoon that would likely result in associated
H$\alpha$ emission).  The total $Q$ for the entire galaxy is
$3\times10^{53}~s^{-1}$ for M33 and $1.5\times10^{53}~s^{-1}$ for
NGC~6946 (Kennicutt 1983 scaled to our adopted distance).  Thus, the
\UDHII\ regions contain 1\% and 7\% of the total ionizing photons.  If
the star formation has been reasonably continuous in these systems, a
plausible estimate for the typical duration of the \UDHII\ phase is
0.01 and 0.07 times the typical \HII\ region lifetime ($\times10^{7}$
yr).  This implies a mean age of less than 1 Myr for \UDHII\ regions.

The third, and perhaps weakest, piece of evidence in favor of the
extreme youth of \UDHII\ regions comes from their high visual
extinctions.  In this article, we note that many of the ionizing star
clusters within the \UDHII\ regions are not visible at optical
wavelengths.  In some cases, only diffuse $H\alpha$ emission is seen.
Sams \etal\ (1994) show that in NGC~253, the extinctions due to dust
reach local maxima, as high as $A_V=15$ mag at the positions of the
radio sources we have identified as \UDHII\ regions.  This picture is
consistent with {\UDHII}s being extremely young \HII\ regions still
hidden from view by the dust associated with their natal molecular
clouds.  However, one must also bear in mind that we cannot rule out
screens of dust not physically associated with the regions of 
optically thick free-free emission.

\subsection{\it Future Work}
We expect that deliberate radio continuum searches will continue to
find \UDHII\ regions in all galaxies with sufficiently high levels of
recent star formation.  It is becoming clear that we are seeing a {\it
continuum} of sizes and luminosities for extragalactic massive star
clusters in the earliest stages of their evolution.  We predict that
the natal cocoons of \UDHII\ regions should be prodigious emitters in
the mid- to far-infrared regimes.  In fact, mid- to far- infrared
observations will allow us to search for an {\it even earlier} stage
of massive star cluster evolution; before the stars have begun
ionizing their surrounding ISM, they should go through a ``hot core''
phase, analogous to individual massive stars in the Galaxy.  This
phase will be defined by extremely dense and warm gas that is not
associated with strong free-free emission.  Therefore, further study
of these sources will be greatly enhanced by upcoming mid- to
far-infrared telescope missions, such as the Space Infrared Telescope
Facility (SIRTF).  Recognition of the ubiquity of the \UDHII\ phase of
massive star formation pushes one step closer to understanding the
genesis mechanisms of all star clusters, from small associations to
giant proto-globular clusters.

\acknowledgments It is a delight to thank Paul Crowther, Dick McCray,
and Sara Beck for useful discussions on this subject.  Ed Churchwell
provided useful feedback on a draft of this paper, for which we are
grateful.  The comments from the anonymous referee led to many
improvements in the manuscript.  We also extend our appreciation to
Soeren Larsen for the use of his optical images in this study.  The
Digitized Sky Surveys were produced at the Space Telescope Science
Institute under U.S. Government grant NAG W-2166. The images of these
surveys are based on photographic data obtained using the Oschin
Schmidt Telescope on Palomar Mountain and the UK Schmidt
Telescope. The plates were processed into the present compressed
digital form with the permission of these institutions.  K.E.J. is
pleased to acknowledge support for this work provided by NASA through
a Graduate Student Researchers Fellowship.  P.S.C. appreciates
continuous support from the National Science Foundation.

\clearpage
\begin{deluxetable}{lllccccc}
\tabletypesize{\scriptsize}
\tablecaption{UD{H\,{\small{\sc II}}} candidates in M33 \label{tbl-1}}
\tablewidth{0pt}
\tablenotetext{a}{These coordinates are accurate to approximately 
the half beam width of $\approx 3 - 4''$.}
\tablehead{
\colhead{\#} &
\colhead{R.A.} & 
\colhead{Dec.} & 
\colhead{$L_6 $} &
\colhead{$L_{20} $}  & 
\colhead{$Q_{Lyc} $} &
\colhead{$\alpha^{20}_{6}$}
\\
&
&
&
\colhead{$\times 10^{23}$} &
\colhead{$\times 10^{23}$} &
\colhead{$\times 10^{49}$} &
&
\\
&
\colhead{(2000)} &
\colhead{(2000)} &
\colhead{$erg/s/Hz$} &
\colhead{$erg/s/Hz$} &
\colhead{$s^{-1}$} &
&
}
\startdata

1&1:33:02.4&30:46:42.9&11.9$\pm$0.8 & 6.8$\pm$1.7&9.7$\pm$0.9 &0.46$\pm$0.11 \\
2&1:33:16.0&30:56:45.9&16.9$\pm$1.7&14.4$\pm$2.5&13.9$\pm$1.6 &0.13$\pm$0.09 \\
3&1:33:16.5&30:52:50.3&55.9$\pm$1.7&35.6$\pm$ 2.5&45.8$\pm$1.4 &0.37$\pm$0.03\\
4&1:33:37.5&30:47:19.3& 9.3$\pm$0.8&7.6$\pm$0.8&7.6$\pm$6.2 &0.17$\pm$0.06 \\
5&1:33:39.2&30:38:06.9& 9.3$\pm$0.8 & 7.6$\pm$1.7&7.6$\pm$6.2 &0.17$\pm$0.10 \\
6&1:33:43.6&30:39:07.1& 8.5$\pm$0.8 & 7.6$\pm$1.7&6.9$\pm$0.8 &0.09$\pm$0.10 \\
7&1:33:48.2&30:39:17.8& 3.4$\pm$0.8 & 2.5$\pm$0.8&2.8$\pm$0.7 &0.25$\pm$0.17 \\
8&1:33:59.8&30:32:45.3& 3.4$\pm$0.8 & 2.5$\pm$0.8&2.8$\pm$0.7 &0.25$\pm$0.17 \\
9&1:34:00.2&30:40:47.7&46.6$\pm$0.8&42.4$\pm$1.7&38.2$\pm$2.2 &0.08$\pm$0.02 \\
10&1:34:02.2&30:38:40.7&28.8$\pm$2.5&26.3$\pm$2.5&23.6$\pm$2.5 &0.07$\pm$0.06\\
11&1:34:06.4&30:41:45.6&11.$\pm$0.8 & 7.6$\pm$0.8&9.0$\pm$0.9 &0.30$\pm$0.06 \\
12&1:34:13.7&30:34:51.4&11.$\pm$0.8 & 7.6$\pm$0.8&9.0$\pm$0.9 &0.30$\pm$0.06 \\
13&1:34:17.3&30:33:43.4&7.6$\pm$0.8 & 6.8$\pm$0.8&6.2$\pm$0.8 &0.09$\pm$0.07 \\
14&1:34:38.9&30:43:59.7&7.6$\pm$0.8 & 5.9$\pm$1.7&6.2$\pm$0.8 &0.21$\pm$0.13 \\

\enddata
\end{deluxetable}

\clearpage
\begin{deluxetable}{lllcccccc}
\tabletypesize{\scriptsize}
\tablecaption{UD{H\,{\small{\sc II}}} candidates in NGC~253 \label{tbl-2}}
\tablewidth{0pt}
\tablenotetext{a}{These coordinates are accurate to approximately 
the half beam width of $\approx 0.1 - 0.2''$.}
\tablehead{
\colhead{\#} &
\colhead{R.A.} & 
\colhead{Dec.} & 
\colhead{$L_{1.3} $} &
\colhead{$L_{2} $}  & 
\colhead{$L_{3.6} $}  & 
\colhead{$Q_{Lyc} $} &
\colhead{$\alpha^{3.6}_{1.3}$} &
\\
&
&
&
\colhead{$\times 10^{24}$} &
\colhead{$\times 10^{24}$} &
\colhead{$\times 10^{24}$} &
\colhead{$\times 10^{49}$} &
&
\\
&
\colhead{(2000)} &
\colhead{(2000)} &
\colhead{$erg/s/Hz$} &
\colhead{$erg/s/Hz$} &
\colhead{$erg/s/Hz$} &
\colhead{$10^{49} s^{-1}$} &
&

}
\startdata
1&0:47:32.75&-25:17:20.94&6.3$\pm$1.3&---&5.$\pm$0.5&60$\pm$13&0.19$\pm0.10$ \\
2&0:47:32.85&-25:17:20.34&33.6$\pm$6.7&25.6$\pm$2.6&20.5$\pm$2.0&319$\pm$66&
0.41$\pm$0.10\\
3&0:47:33.05&-25:17:18.25&59.1$\pm$11.8&58.6$\pm$5.9&49.0$\pm$4.9&561$\pm$120& 0.15$\pm$0.10\\
4&0:47:33.05&-25:17:17.65&20.6$\pm$4.1&---&18.1$\pm$1.8&196$\pm$41&0.11$\pm$0.10 \\
5&0:47:33.25&-25:17:15.55&50.4$\pm$10.1&44.2$\pm$4.4&29.7$\pm$3.&479$\pm$100&0.43$\pm$0.10\\

\enddata
\end{deluxetable}

\clearpage
\begin{deluxetable}{lllccccc}
\tabletypesize{\scriptsize}
\tablecaption{UD{H\,{\small{\sc II}}} candidates in NGC~6946 \label{tbl-3}}
\tablewidth{0pt}
\tablenotetext{*}{These luminosities represent less than a 4$\sigma$ 
detection.}
\tablenotetext{a}{These coordinates are accurate to approximately 
the half beam width of~$\approx 1''$.}
\tablehead{
\colhead{\#} &
\colhead{R.A.} & 
\colhead{Dec.} & 
\colhead{$L_6$} &
\colhead{$L_{20}$}  & 
\colhead{$Q_{Lyc}$} &
\colhead{$\alpha^{20}_{6}$} &
\\
&
&
&
\colhead{$\times 10^{24}$} &
\colhead{$\times 10^{24}$} &
\colhead{$\times 10^{49}$} &
&
\\
&
\colhead{(2000)} &
\colhead{(2000)} &
\colhead{$erg/s/Hz$} &
\colhead{$erg/s/Hz$} &
\colhead{$s^{-1}$} &
&
}
\startdata
1&20:34:19.79&60:10:06.46&11.3$\pm$1.9&10.4$\pm$1.3&92.7$\pm$16&0.07$\pm$0.09\\
2&20:34:22.58&60:10:34.12&43.6$\pm$1.9&33.9$\pm$2.2&358.0$\pm$26&0.21$\pm$0.03 \\
3&20:34:32.35&60:10:12.48&3.8$\pm$0.9&1.3$\pm$0.9*&30.9$\pm$7.9&0.88$\pm$0.32\\
4&20:34:33.89&60:11:25.07&4.7$\pm$1.3&3.8$\pm$0.9&38.6$\pm$11&0.17$\pm$0.16\\
5&20:34:39.70&60:08:22.79&6.0$\pm$0.6&1.9$\pm$0.6*&49.05$\pm$.8&0.94$\pm$0.14\\
6&20:34:49.38&60:08:00.64&4.1$\pm$0.6&3.1$\pm$0.9&33.5$\pm$5.5&0.23$\pm$0.14\\
7&20:34:54.25&60:08:53.62&5.6$\pm$1.3&5.3$\pm$0.9&46.3$\pm$11&0.05$\pm$0.13\\
8&20:34:54.52&60:07:39.73&8.8$\pm$0.9&8.2$\pm$0.9&72.1$\pm$8.7&0.06$\pm$0.07\\
9&20:34:56.49&60:08:20.25&2.5$\pm$0.6&1.9$\pm$0.9*&20.6$\pm$5.3&0.23$\pm$0.23\\
10&20:35:03.67&60:10:59.66&3.4$\pm$0.9&3.1$\pm$1.3&28.3$\pm$7.9&0.08$\pm$0.22\\
11&20:35:04.35&60:09:46.09&3.8$\pm$1.3&2.2$\pm$1.3*&30.9$\pm$10&0.45$\pm$0.30\\
12&20:35:05.08&60:10:57.44&1.3$\pm$0.9*&0.9$\pm$0.6*&10.3$\pm$7.7&0.30$\pm$0.42\\
13&20:35:13.95&60:08:52.14&6.6$\pm$1.3&4.1$\pm$0.9&54.0$\pm$11&0.39$\pm$0.13\\
14&20:35:18.06&60:09:06.17&10.4$\pm$1.3&6.6$\pm$1.9&85.3$\pm$11&0.37$\pm$0.14\\
15&20:35:22.09&60:07:22.90&5.3$\pm$1.3&3.80$\pm$.9&43.8$\pm$11&0.27$\pm$0.15\\
16&20:35:24.12&60:08:42.61&3.8$\pm$0.9&2.2$\pm$0.6*&30.9$\pm$7.9&0.45$\pm$0.15\\
\enddata
\end{deluxetable}


\clearpage

\begin{deluxetable}{llll}
\tabletypesize{\scriptsize}
\tablecaption{Comparison with optical images of M33 \label{M33.tab}}
\tablewidth{0pt}
\tablehead{
\colhead{Source} & 
\colhead{B-band} &
\colhead{H~II} &
\colhead{GDKG} 
\\
\colhead{\#} & 
\colhead{Counterpart?} &
\colhead{Counterpart?} &
\colhead{\#}  

}
\startdata
1 & point source offset? & HBW~193 ? \tablenotetext{a}{HBW refers to the 
Hodge et al. (1999) catalog.} & 15 \\
2 & diffuse, point source offset? &  BCLMP~638 \tablenotetext{b}{BCLMP 
refers to the Boulesteix et al. (1974) catalog.} & 33 \\
3 & diffuse, point source &  BCLMP~623 & 34 \\
4 & no &  BCLMP~611 ? & 67 \\
5 & diffuse emission &  BCLMP~35/36 & 71\\
6 & diffuse emission &  BCLMP~39/40 & 82 \\
7 & no, dust lane? &  BCLMP~43 ? & 91 \\
8 & diffuse emission &  BCLMP~703 ? & 128\\
9 & no, dust lane? & Z~171/179 \tablenotetext{c}{Z refers to the Court\'es
et al. (1987) catalog.} & 129\\
10 & diffuse emission offset? &  BCLMP~87 & 137\\
11 & point source &  BCLMP~77 & 142 \\
12 & point source offset? &  BCLMP~714 & 151\\
13 & diffuse emission &  BCLMP~712 & 161 \\
14 & complex source &  BCLMP~749/750 & 178 \\
\enddata
\tablenotetext{d}{GDKG refers to the reference number in
\citet{gordon99}.}
\end{deluxetable}

\clearpage

\begin{deluxetable}{llll}
\tabletypesize{\scriptsize}
\tablecaption{Comparison with optical images of NGC~253 \label{N253.tab}}
\tablewidth{0pt}
\tablehead{
\colhead{Source} & 
\colhead{I-band} &
\colhead{H$\alpha$} &
\colhead{UV}\\
\colhead{\#} & 
\colhead{Counterpart?} &
\colhead{Counterpart?} &
\colhead{\#}
}
\startdata
1 & complex source &  complex source & 5.45-42.8 \\  
2 & complex source &  complex source & 5.54-42.2 \\
3 & complex source? &  diffuse emission? & 5.72-40.1\\
4 & complex source? &  diffuse emission? & 5.73-39.5\\
5 & diffuse emission?  &  diffuse emission & 5.90-37.4\\
\enddata
\tablenotetext{a}{UV \# refers to the reference number in \citet{ulvestad97}.}
\end{deluxetable}

\clearpage

\begin{deluxetable}{llll}
\tabletypesize{\scriptsize}
\tablecaption{Comparison with optical images of NGC~6946 
\label{N6946.tab}}
\tablewidth{0pt}
\tablehead{
\colhead{Source} & 
\colhead{R-band} &
\colhead{H$\alpha$} &
\colhead{LDG}\\
\colhead{\#} &
\colhead{Counterpart?} &
\colhead{Counterpart?} &
\colhead{\#}
}
\startdata
1 & diffuse emission & --- & 3 \\
2 & no  &  --- & 5 \\
3 & diffuse emission & faint source? & 10 \\
4 & diffuse emission &  diffuse emission & 11 \\
5 & no &  no & 24 \\
6 & point source & point source & 42 \\
7 & no & no & 73 \\
8 & complex source & complex source & 76 \\
9 & no & no & 79 \\
10 &  no  & no & 87 \\
11 & no &  no & 90 \\
12 & diffuse emission & diffuse emission & 94 \\
13 & diffuse emission & diffuse emission & 109 \\
14 & diffuse emission & --- & 111 \\
15 & no & --- & 113 \\
16 & no & --- & 116 \\

\enddata
\tablenotetext{a}{LDG \# refers to the reference number in \citet{lacey97}.}
\end{deluxetable}

\clearpage

\begin{figure}
\figcaption{Locations of the detected UD{H\,{\small{\sc
II}}}s in M33 are shown (along with their number in Table~\ref{tbl-1}) with
respect to the B-band images (shown in gray scale).  The identification
circles are $\sim~7''$ in diameter, reflecting the beam size of the 
radio observations.  North is up and
East is left.  These images are all approximately $1'.5$ on a side. 
\label{M33bband}}
\end{figure}

\begin{figure}
\figcaption{Locations of the detected UD{H\,{\small{\sc II}}}s in
NGC~253 are shown (along with their number in Table~\ref{tbl-2}) with
respect to the H$\alpha$ image (left) and I-band image (right).  The
identification circles are $\sim 2''$ in diameter, reflecting the
astrometric precision of {\it HST}.  North is up and East is left.
The H$\alpha$ image is approximately $12''.7~\times~12''.7$, and the
I-band image is approximately $11''.6~\times~11''.6$. \label{N253}}
\end{figure}


\begin{figure}
\figcaption{Locations of the detected UD{H\,{\small{\sc II}}}s
candidates \#3--12 in NGC~6946 are shown (along with their number in
Table~\ref{tbl-3}) with respect to the H$\alpha$ (left) and R-band
(right) images of \citet{larsen99}.  The identification circles are
$\sim 3''$ in radius, reflecting the relative astrometric uncertainty.
North is up and East is left.  These images are approximately
$1'.6~\times~1'.6$. \label{N6946}}
\end{figure}

\begin{figure}
\figcaption{Locations of the detected
UD{H\,{\small{\sc II}}}s are shown (along with their number in
Table~\ref{tbl-3}) with respect to the Digitized Sky Survey (sources
1--2 and 13--16).  The identification circles are $\sim 3''$ in
radius, reflecting the relative astrometric uncertainty.  North is up
and East is left.  These Digitized Sky Survey images are approximately
$3'.6~\times~3'.6$. \label{N6946dss}}
\end{figure}

\begin{figure}
\plotone{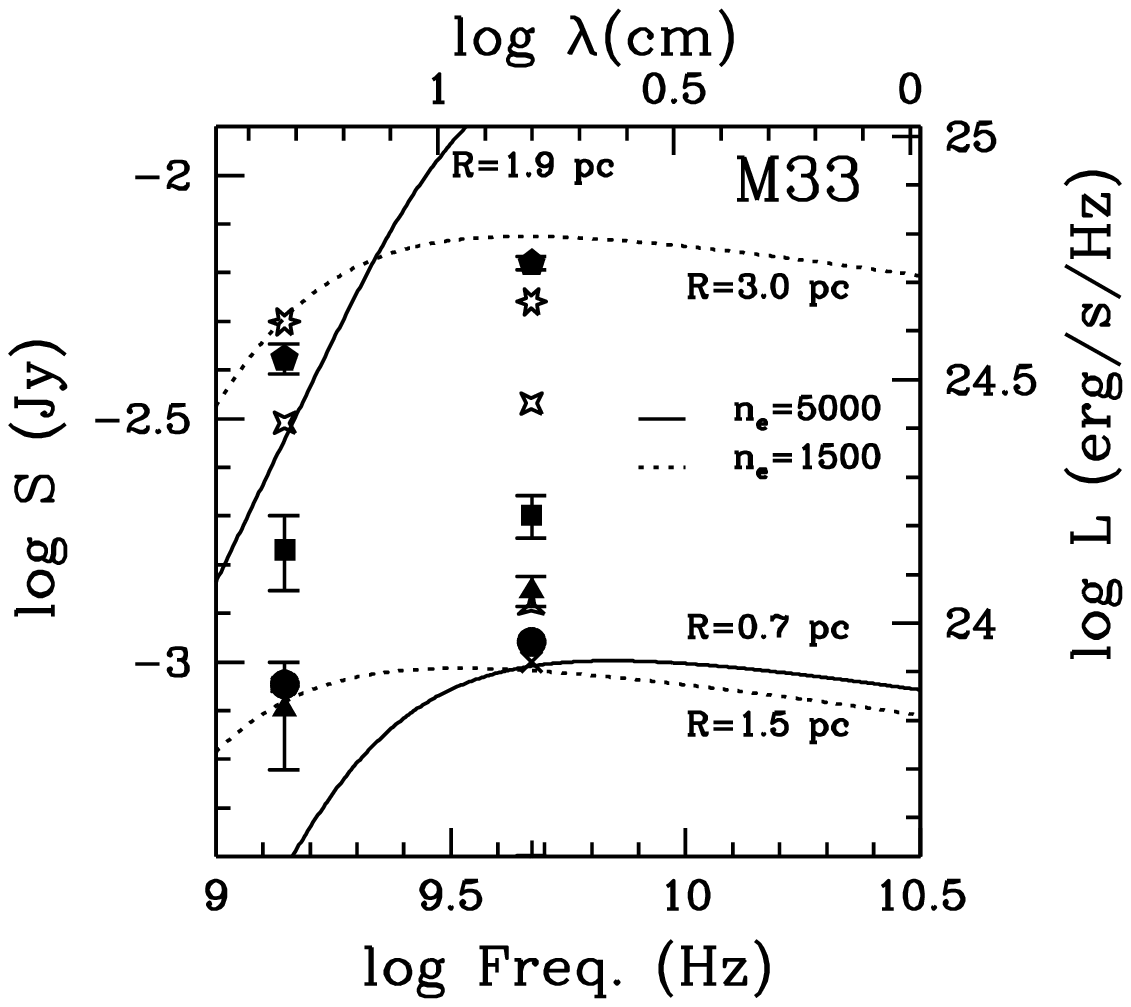}
\figcaption{The radio fluxes and luminosities for the brightest
UD{H\,{\small{\sc II}}}s candidates in M33 from Table~\ref{tbl-1}.  A
different symbol is used for each source.  A typical range of
uncertainties is indicated, and we refer the reader to
Table~\ref{tbl-1} for the remaining values.  The radio data are
consistent with model \HII\ regions (solid and dashed lines) having
electron densities $n_e=1500-5000$ cm$^{-3}$ and radii $R=0.7-3$ pc.
\label{M33.plot}}
\end{figure}

\begin{figure}
\plotone{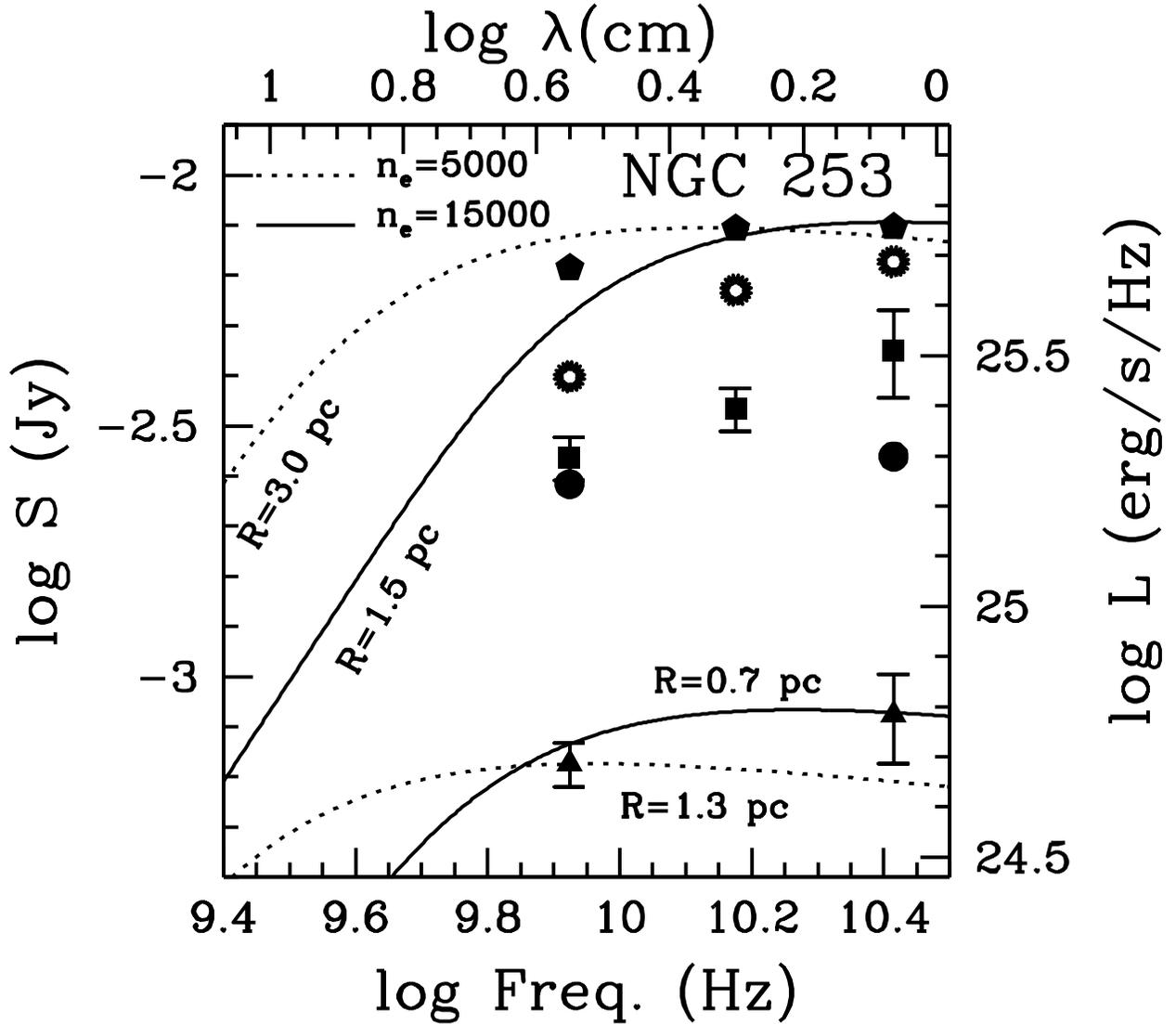}
\figcaption{The radio fluxes and luminosities
for the UD{H\,{\small{\sc II}}}s candidates in NGC~253. A different
symbol is used for each source in Table~\ref{tbl-2}.
A typical range of
uncertainties is indicated, and we refer the reader to
Table~\ref{tbl-2} for the remaining values.
The radio data imply high free-free optical
depths at frequencies as high as 15 GHz.  The data
are consistent with model \HII\ regions (solid and dashed lines)
having electron densities $n_e=5000-15000$ cm$^{-3}$ and
radii $R=0.4-2$ pc.  
\label{N253.plot}}
\end{figure}

\begin{figure}
\plotone{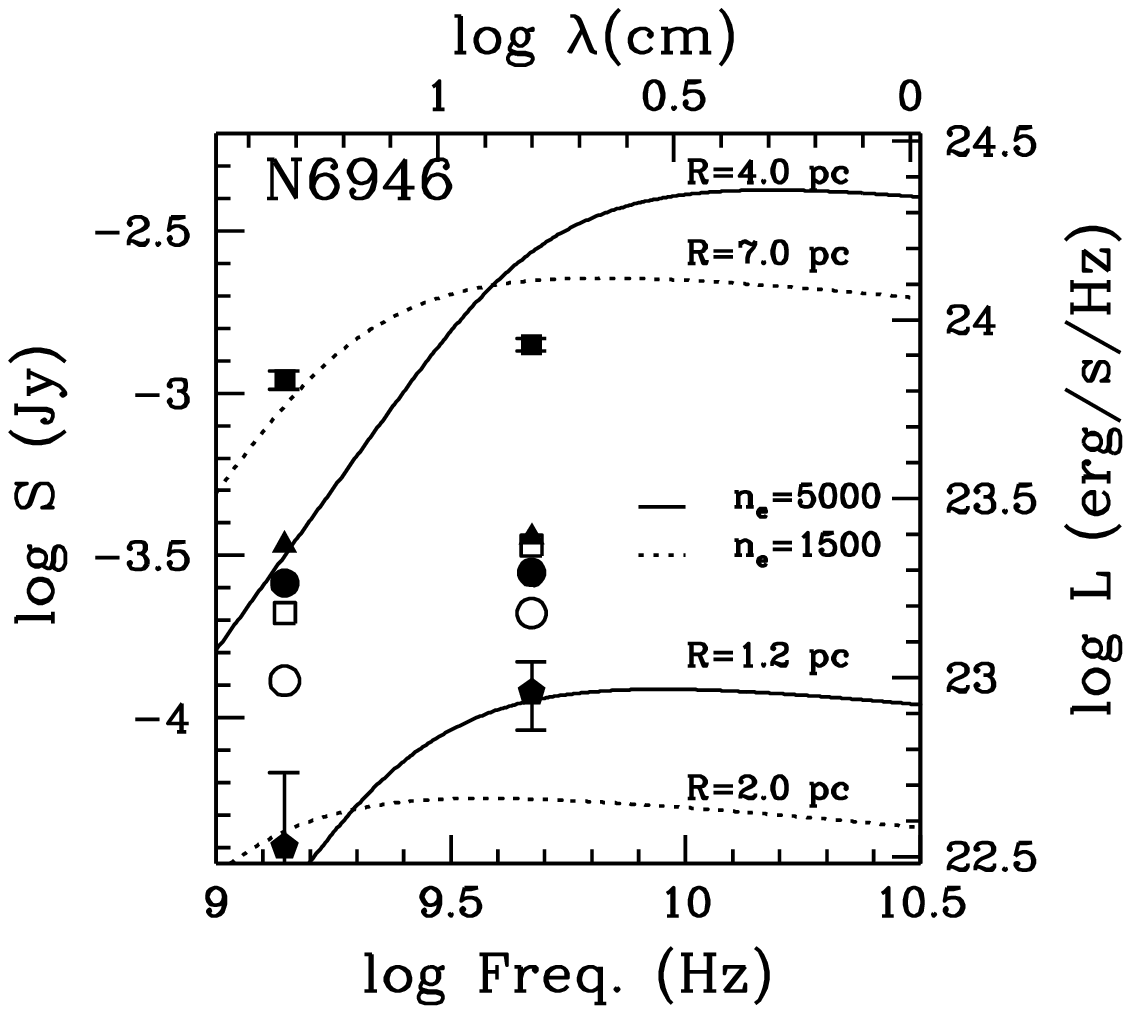}
\figcaption{The radio fluxes and luminosities
for the brightest UD{H\,{\small{\sc II}}}s candidates 
in NGC~6946 in Table~\ref{tbl-3}.  A different
symbol is used for each source.
A typical range of
uncertainties is indicated, and we refer the reader to
Table~\ref{tbl-3} for the remaining values. 
The data are consistent with model \HII\ regions (solid and dashed lines)
having electron densities $n_e=1500-5000$ cm$^{-3}$ and
radii $R=2-7$ pc.   
\label{N6946.plot}}
\end{figure}

\begin{figure}
\plotone{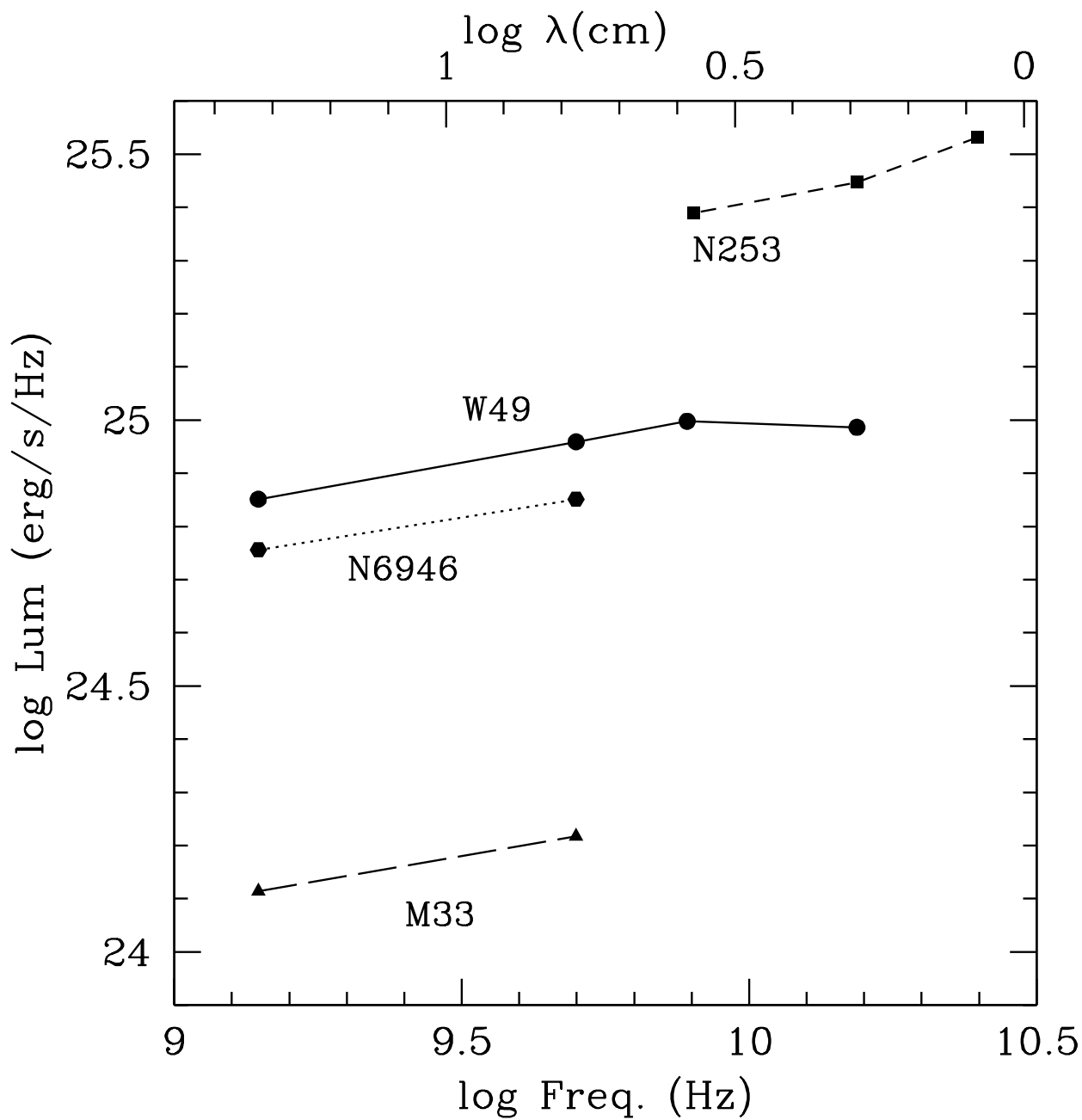}
\figcaption{A comparison of the luminosity and spectral
energy distribution of W49A from \citet{mezger67} and the
{\it mean} luminosity of \UDHII\ regions from M~33, NGC~253, and NGC~6946. 
The luminosity and
spectral energy distribution
of W49A is similar to the \UDHII\ regions in each of the
three galaxies in this study.
\label{W49A.plot}}
\end{figure}

\begin{figure}
\plotone{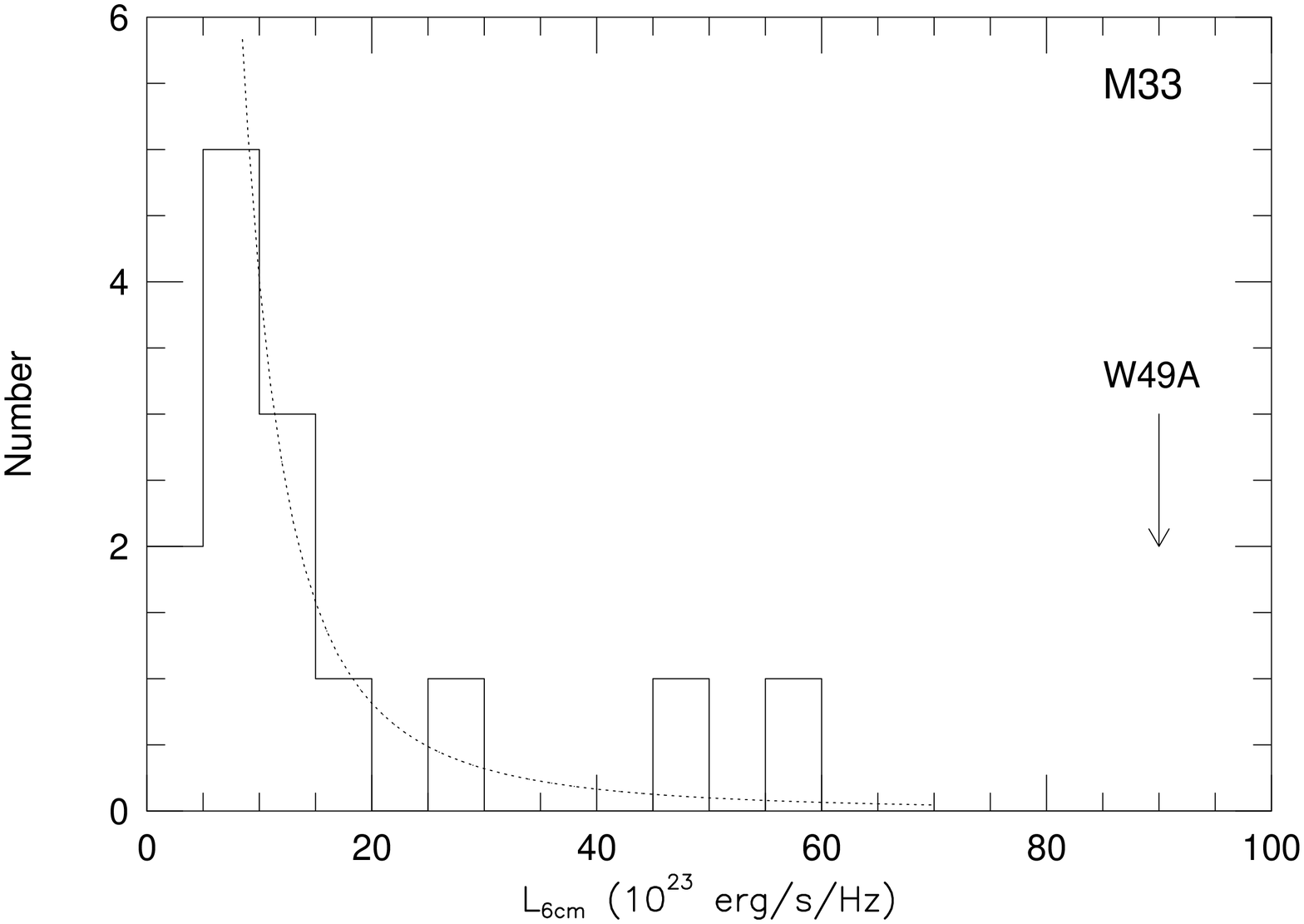}
\end{figure}
\begin{figure}
\plotone{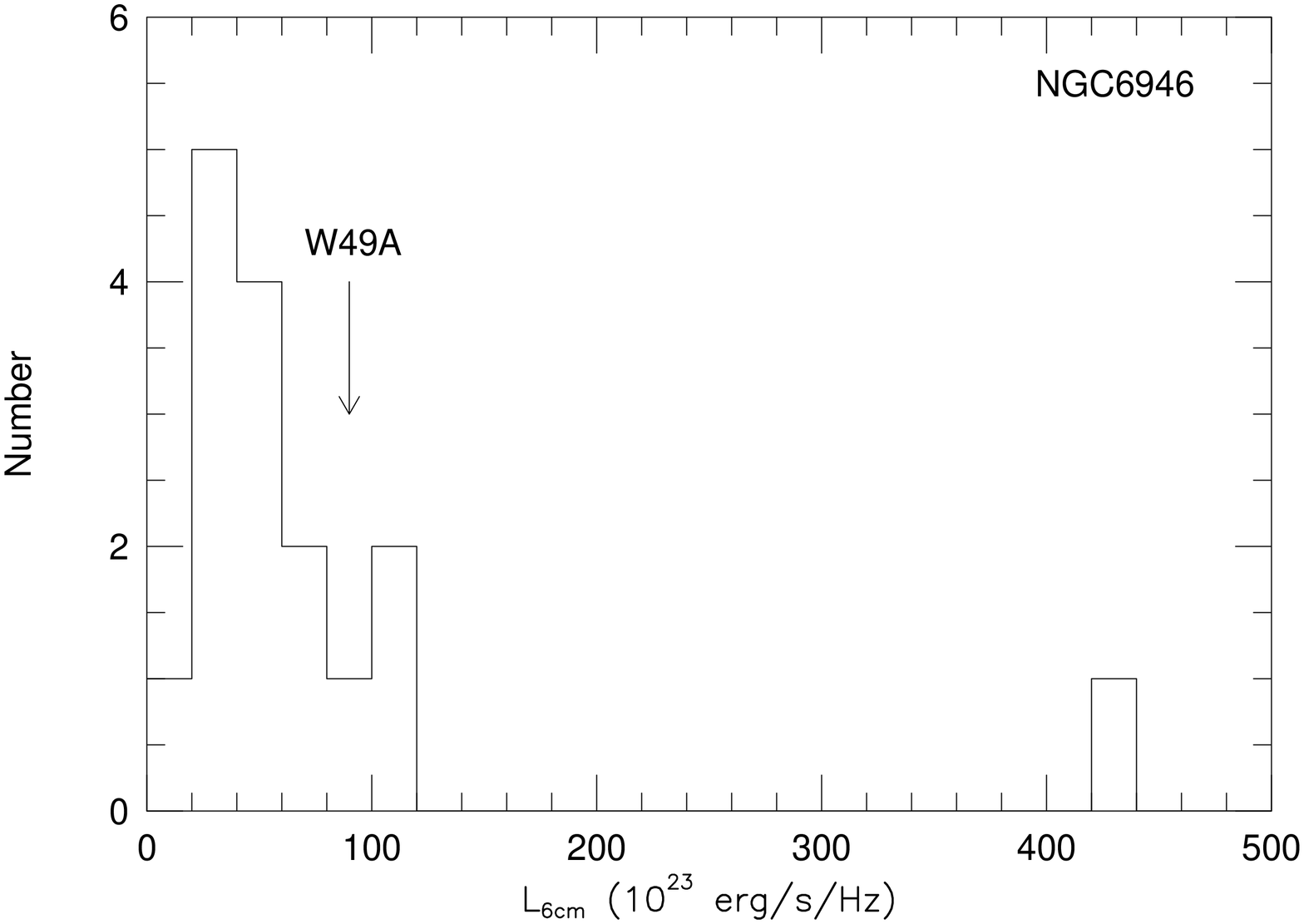}
\end{figure}
\begin{figure}
\plotone{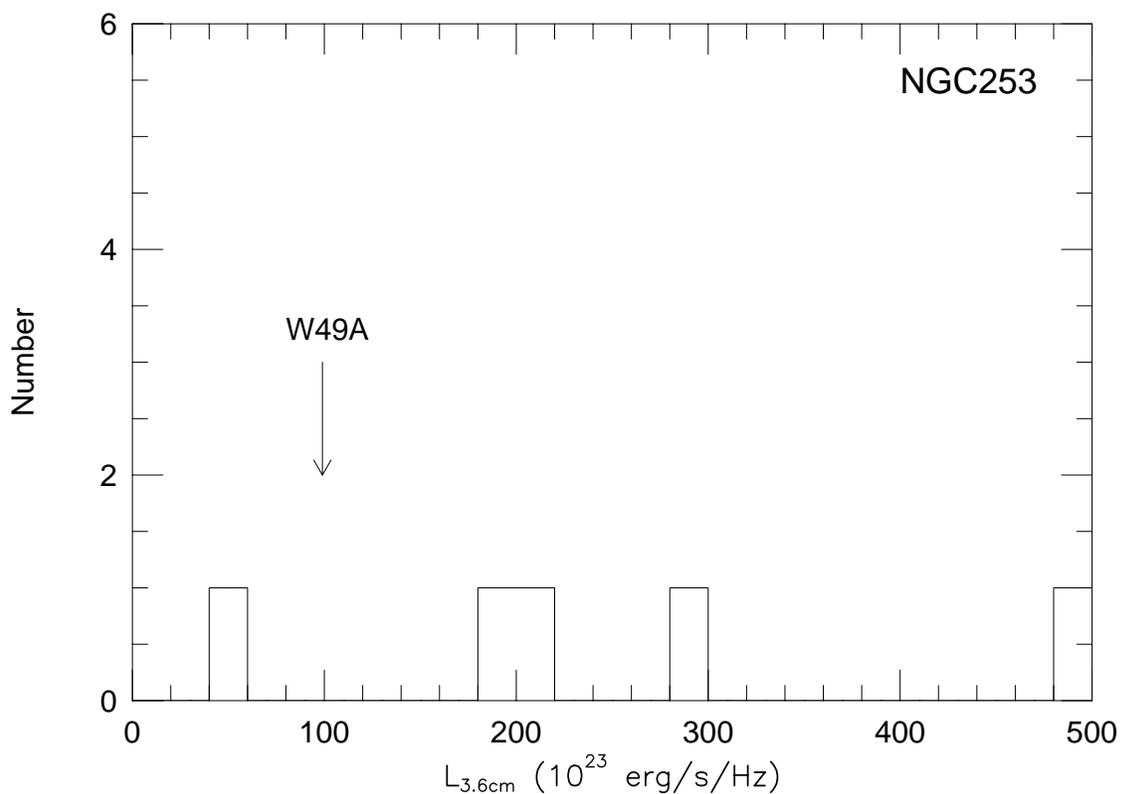}
\figcaption{
Histogram of luminosities for the UD{H\,{\small{\sc II}}} candidates for 
(a) M33, (b) NGC~6946, and (c) NGC~253.  The arrow in each plot 
marks the luminosity of W49A.  
The dotted line in (a) illustrates the luminosity function
of {\it optically selected} \HII\ regions in M33 from \citep{smith89}
that find $N(L) \propto L^{-2.3}dL$.  The completeness limit is
not well determined, but the 5$\sigma$ noise level is $\approx 2 
\times 10^{23}$~erg~s$^{-1}$~Hz$^{-1}$.
\label{histograms}}
\end{figure}


\begin{thebibliography}{}

\bibitem[Afflerbach et al.(1996)]{afflerbach96}
Afflerbach, A., Churchwell, E., Acord, J.M., Hofner, P., Kurtz, S., 
\& Depree, C.G. 1996, ApJS, 106, 423

\bibitem[Beck, Turner, \& Kovo(2000)]{beck00}
Beck, S.C., Turner, J.L., \& Kovo, O. 2000, AJ, 120, 244

\bibitem[Bianchi et al.(2000)]{bianchi00}
Bianchi, S., Davies, J.I., Alton, P.B., Gerin, M. \& Casoli, F. 2000,
A\&A, 353, L13

\bibitem[Boulesteix et al.(1974)]{boulesteix74}
Boulesteix, J., Courtes, G., Laval, A., Monnet, G., \& Petit, H. 
1974, A\&A, 37, 33

\bibitem[Carilli et al.(1992)]{carilli92}
Carilli, C.L., Holdaway, M.A., Ho, P.T.P., \& de Pree, C.G. 1992, 
ApJ, 399, L59

\bibitem[Chandar et al.(1999)]{chandar99}
Chandar, R., Bianchi, L., Ford, H.C., \& Salasnich, B. 1999, PASP, 111, 794

\bibitem[Condon(1992)]{condon92}
Condon, J. J. 1992, ARA\&A, 30, 575

\bibitem[Conti \& Blum(2001)]{conti01}
Conti, P.S. \& Blum, R. 2001, in preparation

\bibitem[Court\'es et al.(1987)]{courtes87}
Court\'es, G., Petit, H., Petit, M., Sivan, J.-P., \& Dodonov, S. 1987, 
A\&A, 174, 28

\bibitem[DePree, Rodr\'iguez, \& Goss(1995)]{depree95}
DePree, C.G., Rodr\'iguez, L.F., \& Goss, W.M. 1995, RMxAA, 31, 39.

\bibitem[DePree, Mehringer, \& Goss(1997)]{depree97}
DePree, C.G., Mehringer, D.M., \& Goss 1997, ApJ, 482, 307

\bibitem[de Vaucouleurs(1979)]{devaucouleurs79}
de Vaucouleurs, G. 1979, ApJ, 227, 729

\bibitem[de Vaucouleurs, de Vaucouleurs, \& Corwin(1976)]{devaucouleurs76}
de Vaucouleurs, G., de Vaucouleurs, A., Corwin, H.G., Jr. 1976,
Second Reference Catalogue of Bright Galaxies (2d ed.; Austin, TX:
Univ. Texas Press) 

\bibitem[Duric et al.(1993)]{duric93}
Duric, N., Viallefond, F., Goss, W.M., \& van der Hulst, J.M. 1993, A\&AS,
99, 217

\bibitem[Elmegreen \& Efremov(1997)]{elmegreen97}
Elmegreen, B.G. \& Efremov, Y.N. 1997, ApJ, 480, 235

\bibitem[Elmegreen, Efremov, \& Larsen(2000)]{elmegreen00}
Elmegreen, B.G., Efremov, Y.N., \& Larsen, S. 2000, ApJ, 535, 748

\bibitem[Freedman, Wilson, \& Madore(1991)]{freedman91}
Freedman, W.L., Wilson, C.D., \& Madore, B.F. 1991, ApJ, 372, 455

\bibitem[Garcia-Gomez \& Athanassoula(1991)]{garcia91}
Garcia-Gomez, C. \& Athanassoula, E. 1991, A\&AS, 89, 159

\bibitem[Gordon et al.(1999)]{gordon99}
Gordon, S.M., Duric, N., Kirshner, R.P., Goss, W.M., \& Viallefond, F.
1999, ApJS, 120, 247

\bibitem[Gwinn, Moran, \& Reid(1992)]{gwinn92}
Gwinn, C.R., Moran, J.M., \& Reid, M.J.1992, ApJ, 292, 149

\bibitem[Hyman et al.(2000)]{hyman00}
Hyman, S.D., Lacey, C.K., Weiler, K.W., \& Van Dyk, S.D. 2000, AJ, 119, 1711

\bibitem[Hodge et al.(1999)]{hodge99}
Hodge, P.W., Balsley, J., Wyder, T.K., \& Skelton, B.P. 1999, PASP, 111, 685

\bibitem[Hollenbach et al.(1994)]{hollenbach94}
Hollenbach, D., Johnstone, D., Lizano, S., \& Shu, F. 1994, ApJ, 428, 654

\bibitem[Kennicutt(1983)]{kennicutt83}
Kennicutt, R.C., Jr. 1983, ApJ, 272, 54

\bibitem[Kennicutt(1984)]{kennicutt84}
Kennicutt, R.C., Jr. 1984, ApJ, 287, 116

\bibitem[Kennicutt(1998)]{kennicutt98}
Kennicutt, R.C., Jr. 1998, ApJ, 498, 541

\bibitem[Keto et al.(1999)]{keto99}
Keto, E., Hora, J.L., Fazio, G.G., Hoffmann, W., \&  Deutsch, L. 1999, 
ApJ, 518, 183

\bibitem[Kobulnicky \& Johnson(1999)]{kj99}
Kobulnicky, H.A. \& Johnson, K.E. 1999, ApJ, 527, 154

\bibitem[Lacey, Duric, \&  Goss(1997)]{lacey97}
Lacey, C., Duric, N., Goss, W.M. 1997, ApJS, 109, 417

\bibitem[Larsen \& Richtler(1999)]{larsen99}
Larsen, S. S. \& Richtler, T. 1999, A\&A, 345, 59

\bibitem[Leitherer et al.(1999)]{leitherer99}
Leitherer, C. et al. 1999, ApJS, 123, 3

\bibitem[Massey et al.(1996)]{massey96}
Massey, P., Bianchi, L., Hutchings, J.B., \& Stecher, T.P. 1996, ApJ, 
469, 629

\bibitem[Mezger, Schraml, \& Terzian(1967)]{mezger67}
Mezger, P.G., Schraml, J., \& Terzian, Y. 1967, ApJ, 150, 807

\bibitem[Mezger \& Henderson(1967)]{mezgerhenderson67}
Mezger, P.G. \& Henderson, A.P. 1967, ApJ, 147, 471

\bibitem[Mohan, Anantharamaiah, \& Goss(2001)]{mohan01}
Mohan, N.R., Anantharamaiah, K.R., \& Goss, W.M. 2001, ApJ in press

\bibitem[Sams et al.(1994)]{sams94}
Sams, B.~J. III, Genzel, R., Eckart, A., Tacconi-Garman, L., \& Hoffman,
R. 1994, ApJ, 430, L33

\bibitem[Sandage \& Tammann(1981)]{sandage81}
Sandage, A. \& Tammann, G.A. 1981, A Revised Shapley-Ames Catalog
of Bright Galaxies (Carnegie Inst. Washington Publ. 635)(Washington:
Carnegie Inst. Washington)

\bibitem[Smith \& Kennicutt(1989)]{smith89}
Smith, T.R. \& Kennicutt, R.C., Jr. 1989, PASP, 101, 649

\bibitem[Smith, Biermann, \& Mezger(1978)]{smith78}
Smith, L.F., Biermann, P., \& Mezger, P.G. 1978, A\&A, 66, 65

\bibitem[Tarchi et al.(2000)]{tarchi00}
Tarchi, A., Neininger, N., Greve, A., Klein, U., Garrington, S.T., 
Muxlow, T.W.B., Pedlar, A., \&  Glendenning, B.E. 2000, A\&A, 358,95

\bibitem[Turner \& Ho(1985)]{turner85}
Turner, J.L. \& Ho, P.T.P. 1985, ApJ, 299, L77

\bibitem[Turner, Ho, \& Beck(1998)]{turner98} 
Turner, J.L., Beck, S.C., \& Ho, P.T.P 1998, ApJ, 532, L109

\bibitem[Turner, Beck, \& Ho(2000)]{turner00}
Turner, J.L., Beck, S.C., \& Ho, P.T.P. 2000, ApJL, 532, 109

\bibitem[Ulvestad \& Antonucci(1997)]{ulvestad97}
Ulvestad, J.S. \& Antonucci, R.R.J. 1997, ApJ, 488, 621

\bibitem[Vacca(1994)]{vacca94}
Vacca, W.D. 1994, ApJ, 421, 140

\bibitem[Watson et al.(1996)]{watson96}
Watson, A.M., Gallagher, J.S., III, Holtzman, J.A., Hester, J.J., 
Mould, J.R., Ballester, G.E., Burrows, C.J., Casertano, S., Clarke, J.T., 
Crisp, D., Evans, R., Griffiths, R.E., Hoessel, J.G., Scowen, P.A., 
Stapelfeldt, K.R., Trauger, J.T., \& Westphtptphal, J.A. 1996, AJ, 112, 534

\bibitem[Weiler \& Panagia(1978)]{weiler78}
Weiler, K.W. \& Panagia, N. 1978, A\&A, 70, 419

\bibitem[Weiler et al.(1986)]{weiler86}
Weiler, K.W., Sramek, R.A., Panagia, N., van der Hulst, J.M., 
\& Salvati, M. 1986, ApJ, 301, 790

\bibitem[Westerhout(1958)]{westerhout58}
Westerhout, G. 1958, Bull. Ast. Inst. Netherlands, 14, 215

\bibitem[Whitmore(2000)]{whitmore00}
Whitmore, B. 2000, in: Celebrating 10 Years of HST Symposium, May 2000,
in press

\bibitem[Wood \& Churchwell(1989a)]{wood89a}
Wood, D.O. \& Churchwell, E. 1989a, ApJS, 69, 831

\bibitem[Wood \& Churchwell(1989b)]{wood89b}
Wood, D.O. \& Churchwell, E. 1989b, ApJ, 340, 265

\end{thebibliography}
\end{document}